\documentstyle[aps,twocolumn,epsf]{revtex}

\begin{document}
\draft

\title{Domain Dynamics of Magnetic Films with Perpendicular Anisotropy}

\author{U.~Nowak}
\address{ 
  Theoretische Tieftemperaturphysik, Gerhard-Mercator-Universit\"{a}t
  Duisburg, 47048 Duisburg, Germany\\
  e-mail: uli@thp.uni-duisburg.de}
\author{J.~Heimel and T.~Kleinefeld}
\address{Angewandte Physik, Gerhard-Mercator-Universit\"at Duisburg,
  47048 Duisburg, Germany}
\author{D.~Weller}
\address{IBM Almaden Research Center, San Jose/CA, USA}

\date{\today}
\maketitle

\begin{abstract}
  We study the magnetic properties of nanoscale magnetic films with
  large perpendicular anisotropy comparing polarization microscopy
  measurements on $\mbox{Co}_{28}\mbox{Pt}_{72}$ alloy samples based on the
  magneto-optical Kerr effect with Monte Carlo simulations of a
  corresponding micromagnetic model. In our model the magnetic film is
  described in terms of single-domain magnetic grains, interacting via
  exchange as well as via dipolar forces. Additionally, the model
  contains an energy barrier which has to be overcome in order to
  reverse a single cell and a coupling to an external magnetic field.
  Disorder is taken into account.

  We focus on the understanding of the dynamics especially the
  temperature and field dependence of the magnetisation reversal
  process.  The experimental and simulational results for hysteresis,
  the reversal mechanism, domain configurations during the reversal,
  and the time dependence of the magnetisation are in very good
  qualitative agreement. The results for the field and temperature
  dependence of the domain wall velocity suggest that for thin films
  the hysteresis can be described as a depinning transition of the
  domain walls rounded by thermal activation for finite temperatures.
\end{abstract}


\section{Introduction}
In recent years great effort was focused on the magnetisation reversal
process in magnetic thin films with perpendicular anisotropy because
of their potential application as high density recording media. In
particular CoPt alloy films were found to be a very promising compound
for magnetic and magneto-optic storage \cite{Weller}.

Two different mechanisms can be thought of to dominate the reversal
process: either nucleation or domain wall motion \cite{Pommier}. Which
of these mechanisms dominates a reversal process depends on the
interplay of the different interaction forces between domains with
different magnetic orientation. In recent experiments on
$\mbox{Co}_{28}\mbox{Pt}_{72}$ alloy films \cite{Theo,Valentin} a
crossover from magnetisation reversal dominated by domain growth to a
reversal dominated by a continuous nucleation of domains was found
depending on the film thickness which was varied from 10nm to 30nm.
Correspondingly, characteristic differences for the hysteresis loops
have been found. Similar results have been achieved by simulations of
a micromagnetic model using zero temperature dynamics
\cite{Theo,Valentin} and Monte Carlo methods respectively
\cite{Nowak}.

It is the goal of this paper to work out the relation between
experiments and simulations. An exact, quantitative description of the
experimental results is hindered by the facts that the observable
length scales in experiment and simulation are different and that
there is no straight forward mapping of the time scale of a Monte
Carlo simulation on experimental time scales.  However, the emphasis
of our work is on a deeper qualitative understanding of the dynamical
aspects of the magnetization reversal and especially on the principal
influence of the field and the temperature on the domain wall
velocity. We find that for thin films these dynamics is governed by a
depinning transition of the domain walls rounded by thermal activation
for finite temperatures.

\section{Experimental Methods}
The CoPt alloy films were prepared under UHV conditions at low
deposition rates on Si substrates covered with a 20nm Pt buffer layer.
In order to achieve a dominant perpendicular anisotropy the films were
grown at rather high substrate temperatures of about $220^{\circ}$C.
The composition of 28 at.\% Co and 72 at.\% Pt is known to exhibit the
maximum of the polar Kerr rotation for the whole range of alloys
\cite{Weller}.  Structural characterization reveals a predominant
fcc(111) texture. The films were usually found to form a
polycrystalline microstructure. Scanning tunneling microscopy was used
to determine the average grain size. The preparation process yields
rather uniform grain size of the order of 20nm almost
independent of the film thicknesses, which cover the range of 5 to
50nm.

X-ray analysis reveals a dispersion of the crystalline c-axis in the
range of about 3 degrees. For the magnetic characterization we applied
magneto-optic Kerr microscopy at very high spatial resolution of about
1$\mu$m \cite{Valentin}. The typical Kerr rotation for the CoPt alloy
films is ranging from $0.1^{\circ}$ to $0.3^{\circ}$. High speed image
processing technique using a low noise CCD camera is able to acquire
domain patterns with millisecond time resolution. The experimental
setup permits to gain complete hysteresis loops with applied external
magnetic field up to 0.5T. An electrical heater assembly provides the
variation of the sample temperature by direct heating using a bifilar
heat wire. Image processing software was developed to determine the
domain wall motion and the average magnetisation of the sample.

\section{Micromagnetic Model}
$\mbox{Co}_{28}\mbox{Pt}_{72}$ alloy films have a polycrystalline
structure. For a theoretical description by a micromagnetic model
\cite{Andra} the film is thought to consist of cells on a square
lattice with a square base of size $L \times L$ where $L = 20$nm. The
height $h$ of the cells is varied from 10nm to 30nm. Due to the large
anisotropy of the CoPt alloy film the cells are thought to be
magnetised perpendicular to the film only with a uniform magnetisation
$M_s$ which is set to the experimental value of $M_s = 365$kA/m for
the saturation magnetisation in these systems \cite{Aachen}. The
cells interact via domain wall energy and dipole interaction. The
coupling of the magnetisation to an external magnetic field $H$ is
taken into account as well as an energy barrier which has to be
overcome during the reversal process of a single cell.

From these considerations it follows that the change of energy caused
by reversal of a cell $i$ with magnetisation $L^2 h M_s \sigma_i$ with
$\sigma_i = \pm 1$ and $\Delta \sigma = \sigma(new)-\sigma(old) = \pm
2$ is:
\begin{eqnarray}
\Delta E_i & = & -\frac{1}{2} L h S_w \Delta \sigma_i
                \sum_{\langle j \rangle} \sigma_j \nonumber\\
           & & + \frac{\mu_0}{4\pi} M_s^2 L h^2 \Delta \sigma_i
                             \sum_j v(\sigma_j,r_{ij}) \nonumber\\
           & & - \mu_0 H L^2 h M_s \Delta \sigma_i
\label{e.ham}
\end{eqnarray}

The first term describes the wall energy $\Delta E_w$ and the sum is
over the four next neighbors. One can expect that the grain
interaction energy density $S_w$ is a reduced Bloch-wall energy
density $S_B$ since the crystalline structure of the system is
interrupted at the grain boundary and also since there may be a chemical
modulation of the CoPt alloy at the grain boundary. We varied the
value of $S_w$ in the simulation and got the best agreement with 
experimental results for a value of $S_w = 0.0022 \mbox{J}/\mbox{m}^2$
which is approximately 50\% of the Bloch-wall energy $S_B$ for this
system \cite{Aachen}.

In the second term describing the dipole coupling $\Delta E_d$ the sum
is over all cells. $r_{ij}$ is the distance between two cells $i$ and
$j$ in units of the lattice constant $L$.  For large distances it is
$v(\sigma_j,r_{ij}) = \frac{\sigma_j}{r_{ij}^3}$. For shorter
distances a more complicated form which is a better approximation for
the shape of the cells which we consider can be determined numerically
and was taken into account.

The third term describes the coupling $\Delta E_H$ to an external
field $H$.

Additionally, an energy barrier $\delta_i$ must be considered which
describes the fact that a certain energy is needed to reverse an
isolated cell. Two reversal mechanisms can be taken into account as
limiting cases: (i) coherent rotation of the magnetisation vector
described by an angel $\theta$: In this case the anisotropy leads to
an energy barrier of $L^2 h K_u$ where $K_u$ is the anisotropy
constant which is $K_u = 200 \mbox{kJ}/\mbox{m}^3$ \cite{Aachen}. (ii)
domain wall motion through the grain: In this case the energy barrier
is $L h S_b$ due to the fact that the domain wall energy is lowered at
the grain boundary. The highest possible value for $S_b$ is the
Bloch-wall energy mentioned above but it can be assumed that the for
the energy barrier relevant value is smaller than the value above
since a Bloch wall is thicker than the size of the cell.  Comparing
these two energies for the reversal mechanism of a CoPt grain one
finds that here domain wall motion through the grains has the lower
energy barrier so that from now on only this mechanism will be taken
into account. We assume that during the reversal process the energy
barrier has its maximum value $L h S_b$ when the domain wall is in the
center of the cell, i.e. when half of the cell is already reversed.
Consequently, the energy barrier which has to be considered in a Monte
Carlo simulation is reduced to $\delta = \max(0, L h S_b - \frac{1}{2}
|(E_w+E_d+E_H)|)$. The simulations are in good agreement with the
experiments using $S_b = 0.0007 \mbox{J}/\mbox{m}^2$.

Note, that we have introduced two different wall energy densities,
$S_w$ for the nearest neighbor interaction, and $S_b$ for the
intrinsic energy barrier. They have the same physical origin but they
have to be handled separately in our model: e.~g., the limit of a
system of isolated grains is described by $S_w \rightarrow 0$ but then
there is still a finite $S_b$ due to the existence of energy barriers
which are relevant for the flip of the isolated cells. The opposite
limit is a system without grain boundaries. Here it is $S_b = 0$,
i.~e., there is no energy barrier but the domain wall has still a
constant, finite energy density $S_w$ which in this limit does not
depend on the position of the wall.

Obviously, the grain sizes in the CoPt alloy are randomly distributed
\cite{Theo}. In order to be realistic, in our simulation disorder has
to be considered (see also \cite{Ruediger} for a discussion of the
role of disorder in a similar simulation). In the model above
this corresponds to a random distribution of $L$ which can hardly be
simulated exactly since it modulates the normalized cell distance
$r_{ij}$ of the dipole interaction. Therefore, as a simplified ansatz
to simulate the influence of disorder we randomly distribute $L$ in
the energy term that describes the coupling to the external field.
Here a random fluctuation of $L$ is most relevant, since this term is
the only one scaling quadratically with $L$. In the simulations we use
a distribution which is Gaussian with width $\Delta$ = 0.1. Through
this kind of disorder our model is mapped on a random-field model.
Other possible origins of disorder in the CoPt films are
e.~g.~fluctuations of the easy axis of the grains or fluctuations of
the saturation magnetisation.  Interestingly these kinds of disorder
would also predominantly influence the last term in Eq. \ref{e.ham}
and hence can be gathered in the random field.

The simulation of the model above was done as in earlier
publications \cite{Nowak,Ruediger} via Monte Carlo methods
\cite{Binder} using the Metropolis algorithm with an additional energy
barrier. Since the algorithm satisfies detailed balance and Glauber
dynamics it allows the investigation of thermal properties as well as
the investigation of the dynamics of the system. Note that in the
limit of low temperatures the Monte Carlo algorithm passes into a
simple energy minimization algorithm with single spin flip dynamics, so
that also the case of zero temperature can be investigated.

The size of the lattice was typically $150 \times 150$. The dipole
interaction was taken into account without any cut-off or mean field
approximation.

\section{Hysteresis and the Reversal Mechanism}
\begin{figure}[t]
  \begin{center}
    \epsfxsize=8cm
    \epsffile{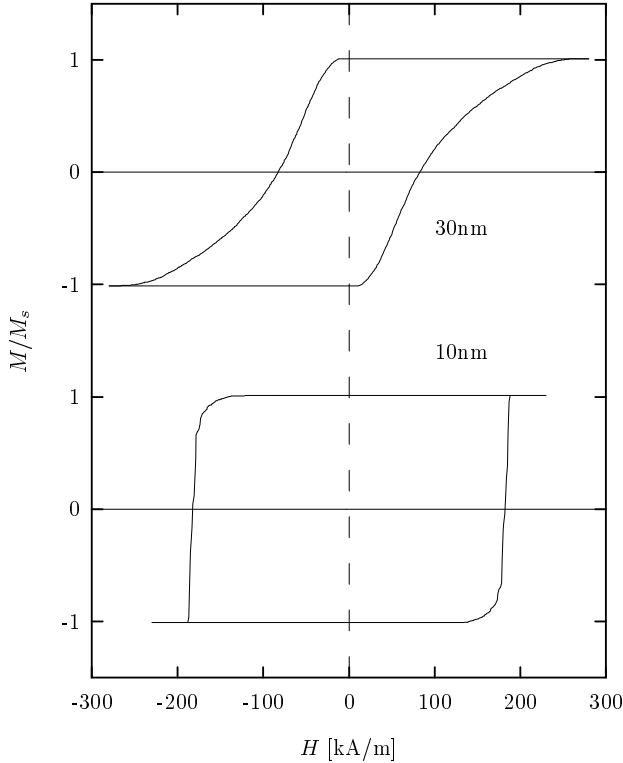}
  \end{center}
  \caption{Simulated hysteresis loops for a 30nm and a 10nm film at $T=300$K.}
  \label{f.hys}
\end{figure}

We start our analysis by comparing the hysteresis loop from
simulations for $T = 300$K, Fig.~\ref{f.hys}, with the corresponding
experimental hysteresis loop for room temperature, see Ref.
\cite{Theo,Valentin}.

\vspace{5mm}

The qualitative agreement is very good. In both cases, experiment as
well as simulation the hysteresis loop of the thin film is nearly
rectangular.  For the thicker film the loop has a finite slope,
especially at the end of the hysteresis.  This can be understood
through the different reversal mechanisms. The reversal of the 10nm
film is dominated by domain wall motion. Once a nucleus begins to grow
the domain wall motion does not stop until the magnetisation has
completely changed.  For the 30nm film the enhanced dipolar forces
stabilize a mixed phase which can be changed only by a further
increase of the external field \cite{Theo,Valentin,Nowak}. The reason
for the enhanced influence of the dipolar forces can be seen in
Eq.\ref{e.ham} where the dipole term is the only term that scales
quadraticly with the film thickness $h$ which means that the influence
of the dipole coupling is neglectable for $h \rightarrow 0$ and on the
other hand strongly growing with increasing film thickness.

Quantitatively, the agreement of the simulation with the experimental
results is reasonable only for the 30nm film while the nucleation
field is much to high for the simulation of the 10nm film. There are
several possible reasons for this effect. First, the nucleation field
which in the case of domain wall motion dominated reversal is the
field where domain wall motion starts depends on the size and on the
shape of the nucleus. In an experimental situation point-like defects
which are not visible in the optical regime can act as nuclei while
there is no such artificial nucleus in our simulation. Second, some of
the parameters of our model like the disorder or the Bloch-wall energy
which influences the prefactors in Eq.\ref{e.ham} may depend on the
film thickness. In our simulation, they are thought to be constant --
for simplicity and since we do not have more information about the
thickness dependence of these parameters.  Third, the time scale at
which a hysteresis loop is observed may play an important role. This
time scale is a few minutes in the experiments and around 60000 MCS
in the simulations. To our knowledge, there is no straight forward way
to map Monte Carlo simulation time on realistic time scales so that
the observation-time windows might be different in experiment and
simulation leading to different nucleation fields. Changing the
sweeping rates changes the nucleation fields due to thermal
activation, experimentally as well as in simulations, but this does not
change the qualitative behavior.

However, it is not the aim of our simulations to calculate the
nucleation field accurately. Rather, the simulations are thought to
contribute to a better understanding of the fundamental properties of
the system, especially to a better understanding of the dynamics.

\section{Dynamics and Temperature Dependence}
\begin{figure}
  \begin{center}
    \epsfxsize=8cm
    \epsffile{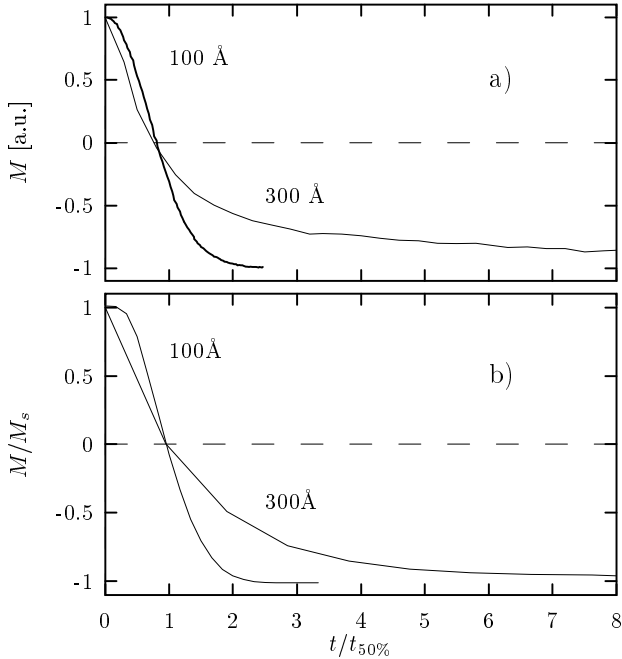}
  \end{center}
  \caption{Magnetisation versus time from measurement, a), after a rapid
    change of the field to 19.33kA/m (10nm) and 13.29kA/m (30nm) and
    from simulations, b), after a rapid change to 260 kA/m (10nm) and
    220 kA/m (30nm).}
  \label{f.dyn}
\end{figure}

The different reversal mechanisms we mentioned in the previous section
manifest themselves also in a change of the dynamical behavior. We
focus on the reversal dynamics, i.~e.~ the time dependence of the
magnetisation after a rapid change of the field to a value which
destabilizes the initial direction of magnetisation. The corresponding
experimental and simulational time dependences of the magnetisation
are compared in Fig.~\ref{f.dyn}. The time-axis is normalized to that
time at which half of the system is reversed.

For the case of the thin films the (absolute) value of the reversed
field is larger than the coercive field while for the thicker film in
the experimental situation a value above the nucleation field but
below the coercive field had to be chosen since for larger fields the
reversal is to fast to be observed with this experimental technique.
Consequently, the long time limit of the corresponding experimental
curve is above -1. Apart from that the experimental results and the
results from simulations agree.

For the case of the 30nm film, i.~e.  for the nucleation driven
reversal there is a rapid change of the magnetisation at the beginning
of the reversal process.  This demonstrates that the reversal process
is dominated by nucleation and not by domain growth processes. In the
limit of a constant nucleation rate $R$ and no growth processes the
change of the magnetisation can be expected to be exponential
\cite{Fatuzzo}.

On the other hand for the case of domain growth the change of
magnetisation is slow at the beginning. If a constant number of nuclei
and a constant domain wall velocity $v$ is assumed, the magnetisation
of the reversed domains grows quadratic in time as long as the domains
are so small that they do not overlap, $M(t) - M_0 \sim -(v t)^2$.
Consequently, the central quantity of the magnetisation reversal
driven by domain wall motion is the domain wall velocity. Therefore,
in the following we will investigate in detail the dependence of the
domain wall velocity on the driving field and the influence of the
temperature. Hence from now on we restrict ourselves to the
investigation of the 10nm film.

\begin{figure}
  \begin{center}
    \epsfxsize=8cm
    \epsffile{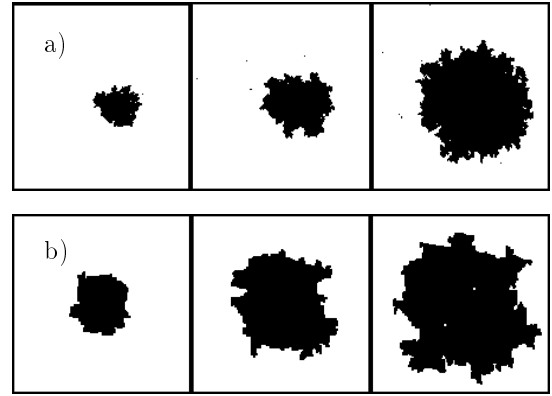}
  \end{center}
  \caption{Domain image from MOKE measurements of a 10nm film
    236, 313, and 468s after a rapid quench to a field of 17.7kA/m,
    a), and simulated domain configurations, of a $150 \times 150$
    system during the reversal of a 10nm film 10, 40, and 60 MCS after
    a rapid quench to a field of 245 kA/m, b)}
  \label{f.dom}
\end{figure}

For the determination of the domain wall velocity within the
simulation we start with a system that has a nucleus of circular shape
with a radius of 19 cells in the center of the $150 \times 150$
system. When we switch on the driving field from the nucleus a domain
starts to grow. For the better observation of the domain growth, in
our flip-algorithm we do not consider cells that are not connected to
the growing domain, i.~e. we exclude the possibility of spontaneous
nucleation. Otherwise we had -- at least for finite temperatures --
the problem that spontaneously new nuclei are build by thermal
activation which with increasing radius overlap with the original
domain.  Fig.~\ref{f.dom} illustrates the time development of a domain
which follows from the method described above and compares it with
experimental domain images. The black regions are reversed domains
following the magnetic field.  The domains look similar although
their size is different (the linear size of the pictures is 3 $\mu$m
in the simulation and 150$\mu$m experimental). This and also a more
detailed analysis \cite{Jost2} leads to the conclusion that the domain
walls are rough and hence self-similar (see \cite{Kardar} for a review
on self similar interfaces). The shape of the domains is - within the
simulations - influenced by the strength of the disorder. Note,
however, that frozen disorder is not the only possible reason for the
roughening of a domain wall. Other possible sources are the dipolar
interactions as well as the dynamics of the model (see e.g.
\cite{Lyberatos} for a discussion of the shape of domains in a model
without disorder). 

From both, the domain configurations from simulations and from
experimental domain images the mean radius $r$ of the domains can be
determined through the area $F$ of the reversed domain as $r =
\sqrt{F/\pi}$, assuming that the domain has a circular shape.
Additionally, the domain radius of the experimental domain images has
been determined as the mean distance of the domain boundary from the
center of the domain. Both definitions of $r$ lead to the same
results.

\begin{figure}
  \begin{center}
    \epsfxsize=8cm
    \epsffile{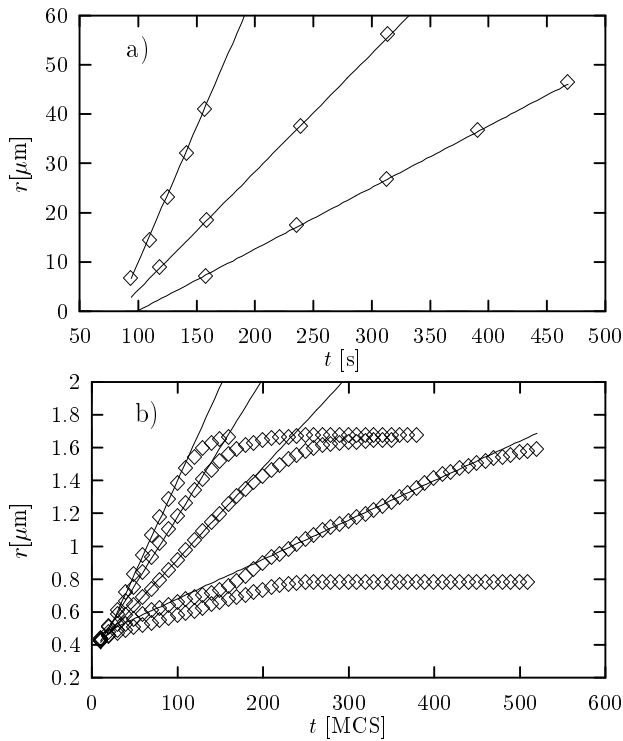}
  \end{center}
  \caption{Measured radius of a domain versus time for $H = 17.7, 18.0, 18.8$
    kA/m. $T = 299$K, a), and radius of the domain versus time from
    simulations for $H = 225, 230, 235, 240, 245$ and zero
    temperature. The solid lines are best fitted.}
  \label{f.kt}
\end{figure}

Fig.~\ref{f.kt}a shows the time dependence of the radius measured at
$T=299$K after a rapid change of the field to $H = 17.7, 18.0$, and
$18.8$kA/m. Obviously, the domain wall velocity is constant and from
the slope of the curves $v$ can be determined.

In order to get a deeper understanding of the influence of the
temperature on the dynamics apart from the ``experimental
temperatures'' ($T \approx 300$K) we also simulated ``extreme
temperatures'' ($T=0$K and $T=600$K).  Fig.~\ref{f.kt}b shows - as
an example - the $r(t)$ behavior from the simulations for $T=0$K and
different fields. For the lowest field shown the domain wall is
pinned, i.~e.  after a short period of rearrangement of the domain
wall its movement stops and the radius remains constant.  The pinning
of the domain wall is due to energy barriers which follow from the
disorder, the dipole field, and the intrinsic energy barrier of the
single cell. For finite temperatures, the domain wall velocity is
always finite (see discussion below). 

\begin{figure}
  \begin{center}
    \epsfxsize=8cm
    \epsffile{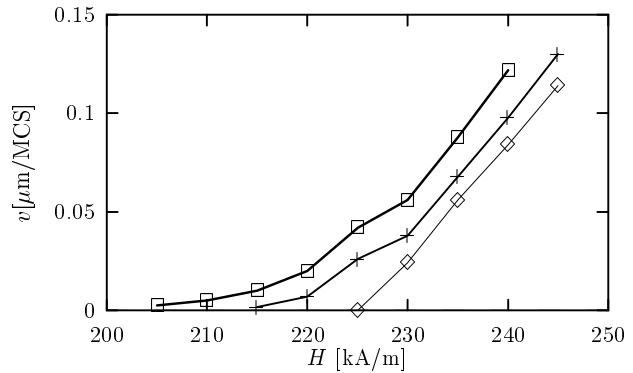}
  \end{center}
  \caption{Simulated domain wall velocity versus driving field for $T = 600,
    300$, and 0K (from above). The solid lines are guides to the eye.}
  \label{f.vh}
\end{figure}

For $r > 150 L/2 = 1.5 \mu \mbox{m}$ the domain reaches the boundary
of the system and - consequently - $r(t)$ saturates.  For smaller $r$
the slope of the $r(t)$ curve is approximately constant and $v$ can be
determined by fitting to a straight line.  Fig.~\ref{f.vh} shows the
resulting dependence of the domain wall velocity on the driving field
for $T=0, 300$, and 600K.

For zero temperature there is a sharp depinning transition
\cite{Kardar} at a critical field $H_c$ from a pinned phase with $v =
0$ to a phase with finite domain wall velocity.  This transition can
be interpreted in terms of a dynamic phase transition with $v \sim
(H-H_c)^\theta$ for $H > H_c$ where in our case the critical exponent
is $\theta \approx 1$, a value which is the mean field result for a
moving elastic interface \cite{Leschhorn} in a random-field.  Also,
this value has been observed in simulations of a soft spin model with
random-fields \cite{Jost}. Hence, it seems to be reasonable to
consider the depinning transition we found to be in the universality
class of a driven interface in the random-field Ising model. This fact
is further confirmed by a dynamical scaling analysis of the structure
of the domain walls in CoPt \cite{Jost2}. The central quantity in this
analysis is the time dependent roughness of the domain wall which -
following the theory of driven interfaces \cite{Kardar} - is
characterized by a certain set of critical exponents. For the case of
the CoPt alloy these exponents are also the same as those
characterizing a driven interface in a random-field Ising model.

Note that in the models mentioned above the only origin for the
depinning transition is the disorder. Here, the value of $H_c$
without disorder is zero. In our model, $H_c$ depends additionally on
the dipole field and the intrinsic energy barrier of the single cell.
A detailed analysis of these dependencies must be left for the future.

\begin{figure}
  \begin{center}
    \epsfxsize=8cm
    \epsffile{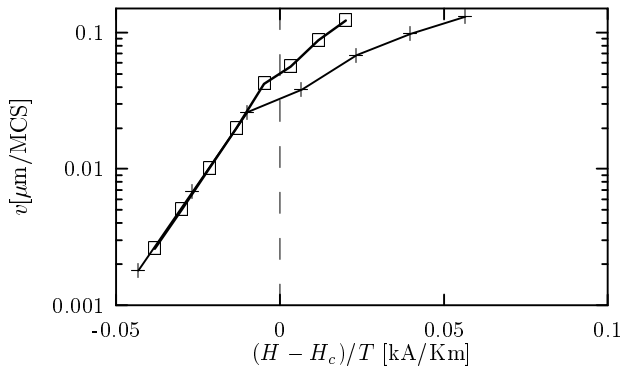}
  \end{center}
  \caption{Scaling plot from Fig.~\ref{f.vh}.}
  \label{f.skal}
\end{figure}

For finite temperatures the transition is smeared since for finite
temperatures there is even for $H<H_c$ for each energy barrier a
finite probability that the barrier can be overcome by thermal
fluctuations. Hence, for finite temperatures we expect a crossover
from the dynamics of the zero temperature depinning transition
explained above to a dynamics dominated by thermal activation where
the corresponding waiting times are exponentially large. For $H < H_c$
the domain wall velocity should decrease like $\ln v \sim (H-H_c)/T$.
To illustrate this in Fig.\ref{f.skal} we show the corresponding
semi-logarithmic scaling plot.  As we expect, the data for the two
different finite temperatures collapse for $H<H_c$ on a straight line.
For $H>H_c$ thermal activation is obviously less relevant since for
these values of the driving fields even for zero temperature the
domain walls move. Consequently, also for finite temperature but
$H>H_c$ the domain wall dynamics is dominated by the zero temperature
depinning transition, i.e. $v \sim H-H_c$.

\vspace{5mm}
\begin{figure}
  \begin{center}
    \epsfxsize=8cm
    \epsffile{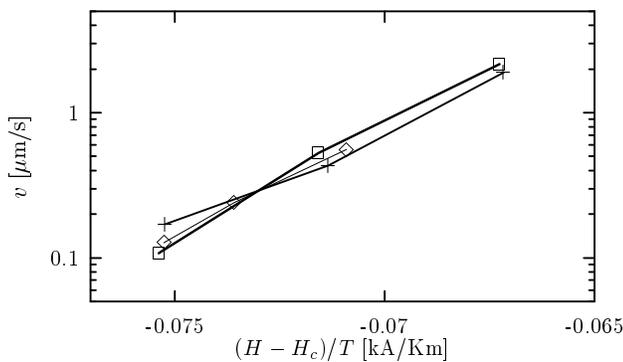}
  \end{center}
  \caption{Scaling plot for the measured domain wall velocities at
    $T=299,335,$ and $394$K.}
  \label{f.eskal}
\end{figure}

Fig.~\ref{f.eskal} shows the same scaling plot for the experimental
data. Obviously, all data are in the regime $H \ll H_c$, so that the
scaling for thermal activation works. We are far away from the regime
$H \approx H_c$ where the crossover to zero temperature depinning
dynamics would start since from Fig.~\ref{f.eskal} it follows $H_c
\approx 40$kA/m. There is no possibility for measurements at higher
fields since for higher fields new domains build up spontaneously and
overlap each other so that it is not possible to follow one single 
domain for the analysis of the domain wall velocity.

\section{Conclusions}
By comparing microscopy measurements on $\mbox{Co}_{28}\mbox{Pt}_{72}$
alloy samples based on the magneto-optical Kerr effect with Monte
Carlo simulations we show that a micromagnetic model can be used for
the understanding of magnetic properties of nanoscale magnetic films
with high perpendicular anisotropy. The experimental and simulational
results of the hysteresis, the reversal mechanism, the domain
configurations during the reversal, the time dependence of the
magnetisation and the temperature and field dependence of the domain
wall velocity are in very good qualitative agreement.

For thin films the reversal is dominated by a growth of domains the
dynamics of which can be described by the domain wall velocity. The
results for the domain wall velocity suggest that for zero temperature
the hysteresis can be understood as a depinning transition of the
domain walls. For finite temperatures the transition is rounded by 
thermal activation and for fields smaller than the depinning field the
domain wall movement is dominated by thermal activation.

\acknowledgments
Work supported by the Deutsche Forschungsgemeinschaft through
Sonderforschungsbereich 166

\end{document}